\documentclass[twocolumn,prl,showpacs,amsfonts,amsmath,assymb,eufrak]{revtex4}
\usepackage{amsmath,amsfonts,amssymb,mathtools,upgreek}
\usepackage{epsfig}

\def\eq#1\en{\begin{equation}#1\end{equation}}  
\def\eqa#1\ena{\begin{align}#1\end{align}}
\def\eqg#1\eng{\begin{gather}#1\end{gather}}
\newcommand{\lb}[1]{\label{e:#1}}
\newcommand{\rlb}[1]{\eqref{e:#1}} 
\newcommand{\nl}{\notag\\}

\newcommand{\sbkt}[1]{\langle#1\rangle}

\newcommand{\sumtwo}[2]%
{\mathop{\sum_{#1}}_{#2}}
\newcommand{\sumthree}[3]%
{\mathop{\mathop{\sum_{#1}}_{#2}}_{#3}}
\newcommand{\sumfour}[4]%
{\mathop{\mathop{\mathop{\sum_{#1}}_{#2}}_{#3}}_{#4}} 
\newcommand{\prodtwo}[2]%
{\mathop{\prod_{#1}}_{#2}}
\newcommand{\mintwo}[2]%
{\mathop{\min_{#1}}_{#2}}
\newcommand{\maxtwo}[2]%
{\mathop{\max_{#1}}_{#2}}
\newcommand{\maxthree}[3]%
{\mathop{\mathop{\max_{#1}}_{#2}}_{#3}}
\newcommand{\limtwo}[2]%
{\mathop{\lim_{#1}}_{#2}}
\newcommand{\suptwo}[2]%
{\mathop{\sup_{#1}}_{#2}}
\newcommand{\supthree}[3]%
{\mathop{\mathop{\sup_{#1}}_{#2}}_{#3}}
\newcommand{\supfour}[4]%
{\mathop{\mathop{\mathop{\sup_{#1}}_{#2}}_{#3}}_{#4}} 
\newcommand{\inftwo}[2]%
{\mathop{\inf_{#1}}_{#2}}
\newcommand{\infthree}[3]%
{\mathop{\mathop{\inf_{#1}}_{#2}}_{#3}}
\newcommand{\inffour}[4]%
{\mathop{\mathop{\mathop{\inf_{#1}}_{#2}}_{#3}}_{#4}} 
\newcommand{\bsr}{\boldsymbol{r}}

\newcommand{\bsv}{\boldsymbol{v}}

\newcommand{\bsR}{\boldsymbol{R}}

\newcommand{\bsV}{\boldsymbol{V}}
\newcommand{\bbC}{\mathbb{C}}

\newcommand{\bbR}{\mathbb{R}}

\newcommand{\ep}{\epsilon}
\newcommand{\up}{\uparrow}
\newcommand{\dn}{\downarrow}

\newcommand{\ph}{\varphi}
\newcommand{\tJ}{\tilde{J}}
\newcommand{\bh}[1]{\sbkt{#1}_{\beta,h}}

\newcommand{\para}[1]{\medskip\par{\em #1}\/.---}
\newcommand{\prop}[1]{{\em #1}\/.---}
\begin{document}
\title{Hohenberg-Mermin-Wagner type theorems for equilibrium models of flocking}

\author{Hal Tasaki}
\affiliation{Department of Physics, Gakushuin University, Mejiro, Toshima-ku, Tokyo 171-8588, Japan}

\date{\today}

%%%%%%%%%%%%%%%%%
\begin{abstract}
We study a class of two-dimensional models of classical hard-core particles with Vicsek-type ``exchange interaction'' that aligns the directions of motion of nearby particles.
By extending the Hohenberg-Mermin-Wagner theorem for the absence of spontaneous magnetization and the McBryan-Spencer bound for correlation functions, we prove that the models do not spontaneously break the rotational symmetry in their equilibrium states at any nonzero temperature.
We thus conclude that the mobility of particles alone does not account for  the  spontaneous symmetry breaking in Vicsek type models.
The origin of the symmetry breaking must be sought in the absence of detailed balance condition, or, equivalently, in the nonequilibrium nature.

There is a 18.5 minutes video on YouTube in which I discuss the background, motivation, and main results of the present paper: \url{https://youtu.be/CD0WEkNl9XA}
\end{abstract}

%05.20.−y Classical statistical mechanics
%05.70.Fh Phase transitions: general studies
%02.30.−f Function theory, analysis

\maketitle
%%%%%%%%%%%%%%%%%%%%%%%%%%%%%%%%%%%%%
%%%%%%%%%%%%%%%%%%%%%%%%%%%%%%%%%%%%%

\para{Introduction}%
Idealized theoretical models of flocking, the formation of clusters of collectively moving self-propelled elements (such as birds), have recently been attracting considerable interest from physics community \cite{SR-REVIEW,VICSEK-REVIEW,MARCHETTI,Ginelli,ChateMahault,Chate}.
Such models are important not only because they shed light on the biological nature of flocking, but also because they lead to novel universality classes in statistical physics.
A prototypical model, known as the Vicsek model, in which nearby self-propelled elements tend to align the directions of motion with each other, was studied numerically in the pioneering work of Vicsek, Czir\'ok, Ben-Jacob, Cohen, and Shochet \cite{VICSEK}.
It was soon realized through intensive studies of the corresponding continuum dynamical model  by Toner, Tu, and Ramaswamy \cite{TONER,TONER2,TONER3,TTR-REVIEW}  that Vicsek-type models may exhibit spontaneous breaking of the rotational symmetry in two or higher dimensions.
See Fig.~6b of \cite{ChateMahault} or Fig.~4a of \cite{Chate} for a definitive numerical evidence for the existence of spontaneous symmetry breaking in the Vicsek model.
The ordered phases were observed experimentally  in biological systems (with nematic order) \cite{NishiguchiNagaiChateSano2017,Tanida} and found numerically in granular systems (with ferromagnetic order)  \cite{Deseigne,Weber,Kumar,Soni}.
There are almost no mathematically rigorous results concerning the ordered phases in rotationally symmetric models of flocking.
See \cite{Kourbane-Houssene} for interesting rigorous results in different active matter systems.

The spontaneous breakdown of the rotational symmetry in a two-dimensional model is in an apparent contradiction with the well known fact, proved first by Hohenberg for quantum particle systems \cite{H} and by Mermin and Wagner for quantum spin systems \cite{MW}, that a two-dimensional system in thermal equilibrium does not spontaneously break continuous symmetry.
This fact has been proven in various models of classical and quantum statistical mechanics.
See, e.g., \cite{MS,FP,TasakiBook}.

There is of course no true contradiction here since these theorems do not apply to the steady state of the Vicsek model, which is not a thermal equilibrium state.
Nevertheless it is natural to ask which physical mechanism is relevant for the violation of the Hohenberg-Mermin-Wagner type theorems.
A common informal explanation is that the motion of self-propelled elements generates effectively long-ranged interaction between the directions of motion of particles, thus violating an essential condition for Hohenberg-Mermin-Wagner type theorems.
In a more sophisticated discussion one focuses on the coupling between the fluctuation of the order parameter (i.e., the direction of the collective motion) and the macroscopic flow of the elements.
See section~3.3 of \cite{Ginelli}.

Such arguments lead us to ask whether a Vicsek type model with detailed balance dynamics can exhibit spontaneous breaking of the rotational symmetry in its steady state, namely, the thermal equilibrium state.
In fact some Hamiltonian models of flocking were studied, and it was reported that behaviors similar to the Vicsek model were observed  \cite{HF1,HF2,HF3}.

In the present paper we study a class of systems of hard-core particles in two dimensions with Vicsek-type ``exchange interaction'' that aligns the directions of motion of nearby particles.
By proving analogues of the Hohenberg-Mermin-Wagner theorem \cite{H,MW} and the McBryan-Spencer bound for correlations \cite{MS}, we rigorously establish that the models do not break the rotational symmetry of the velocities in their equilibrium states.
Note that these equilibrium states are realized as the unique stationary states of the Vicsek like dynamics in which particles move according to Newtonian mechanics, while their velocities are varied stochastically from time to time in such a manner that detailed balance condition holds \cite{GH}.
We thus conclude that the mobility of particles in Vicsek type models is not sufficient to explain the emergence of spontaneous symmetry breaking.

It should be clear that the original proof of Mermin and Wagner \cite{MW}, which relies on the Fourier transformation on the regular lattice, cannot be extended to our models.
We here make use of the method of complex translation introduced by McBryan and Spencer \cite{MS}, which allows us to cover a wide range of models.
Our theorems readily extend to Hamiltonian flock models with extra ``spins'' as in \cite{HF1,HF2,HF3} provided that the particle-particle interaction has a hard-core.

\para{The model and main results}%
We study a classical system of $N$ identical particles in the square region $[0,L]^2$ with periodic boundary conditions.
For $\bsr,\bsr'\in[0,L]^2$, we denote by $|\bsr-\bsr'|$ the Euclidean distance that takes into account the boundary conditions.
We denote the  positions and velocities of the particles as $\bsr_j\in[0,L]^2$ and $\bsv_j\in\bbR^2$, respectively, where $j=1,\ldots,N$ is the label for particles.
Our model is described by the Hamiltonian $H=H_{\rm p}+H_{\rm v}$.
The Hamiltonian for particles is standard, and is given by
\eq
H_{\rm p}=\sum_{j=1}^N\ep(|\bsv_j|)+\mathop{\sum_{j,k=1}^N}_{(j<k)}u(\bsr_j,\bsr_k),
\lb{Hp}
\en
where $\ep(v)$ is an arbitrary one-particle kinetic energy.
One usually sets $\ep(v)=mv^2/2$, but can take a function that has a sharp minimum at certain $v_0$ to mimic the constant speed setting in the Vicsek model.
The two-body potential $u(\bsr,\bsr')$ satisfies the hard-core condition, $u(\bsr,\bsr')=\infty$ if $|\bsr-\bsr'|<a_0$, and is arbitrary otherwise.
We only consider particle number $N$ such that configurations with $\sum_{j<k}u(\bsr_j,\bsr_k)<\infty$ exist.
The exotic Hamiltonian that depends on the directions of the velocities is given by
\eq
H_{\rm v}=-\mathop{\sum_{j,k=1}^N}_{(j<k)}J(\bsr_j,\bsr_k)\frac{\bsv_j}{|\bsv_j|}\cdot\frac{\bsv_k}{|\bsv_k|}
-h\sum_{j=1}^N\frac{v^{\rm x}_j}{|\bsv_j|}.
\lb{Hv}
\en
The first term represents the Vicsek-type  ``exchange interaction''.
We assume that $|J(\bsr,\bsr')|\le J_0$, and $J(\bsr,\bsr')=0$ if $|\bsr-\bsr'|>a_1$.
The second term in \rlb{Hv} is included to test for possible spontaneous symmetry breaking of the rotational symmetry, and $h\ge0$ is the symmetry breaking field.
One has $h=0$ in the standard setting.
The constants $J_0$, $a_0$, and $a_1$ (where we assume $a_0<a_1$) are fixed throughout the paper.

The equilibrium state of the model at  inverse temperature $\beta>0$ is described by the expectation
\eq
\sbkt{\cdots}_{\beta,h}=Z_{\beta,h}^{-1}\int d\bsR\,d\bsV(\cdots)\,e^{-\beta H},
\lb{ex}
\en
where the partition function $Z_{\beta,h}$ is determined from the normalization condition $\sbkt{1}_{\beta,h}=1$.
We wrote $d\bsR=\prod_{j=1}^Nd^2\bsr_j$ and  $d\bsV=\prod_{j=1}^Nd^2\bsv_j$. 

Although the original problem (of classical dynamics) is not rotationally invariant because of the geometry of the region and possible anisotropy in $u$ and $J$, the equilibrium expectation \rlb{ex} is completely invariant under a uniform rotation of all the velocities.
This is a peculiar feature of classical equilibrium statistical mechanics.
We are interested in possible spontaneous breaking of this rotational symmetry.

Our first result is the following extension of the Hohenberg-Mermin-Wagner theorem.

\prop{Theorem 1}%
For any $0<\beta<\infty$ one has
\eq
\lim_{h\dn0}\lim_{L\up\infty}\frac{1}{N}\sum_{j=1}^N\sbkt{v^{\rm x}_j}_{\beta,h}=0,
\lb{HMW}
\en
where the particle number $N$ may depend in an arbitrary manner on the system size $L$ (although it is most natural to fix $N/L^2$ constant).

Since the symmetry breaking field $h>0$ forces $\sbkt{v^{\rm x}_j}_{\beta,h}$ to be positive, \rlb{HMW} establishes that the equilibrium state does not spontaneously break the rotational symmetry.
Recall that the order of the limits in \rlb{HMW} is essential; one trivially has $\lim_{h\dn0}\sbkt{v^{\rm x}_j}_{\beta,h}=0$ for any finite $L$ by continuity.

Let us turn to a more standard setting with $h=0$, and  define the correlation function for the directions of the velocities of two particles by the conditional expectation
\eq
C_\ell(\beta)=
\frac{\bigl\langle\frac{\bsv_j}{|\bsv_j|}\cdot\frac{\bsv_k}{|\bsv_k|}\,\chi^{\ell}_{j,k}\bigr\rangle_{\beta,0}}{\sbkt{\chi^{\ell}_{j,k}}_{\beta,0}},
\lb{C}
\en
for any $j\ne k$, 
where the characteristic function
\eq
\chi_{j,k}^\ell=\begin{cases}
1&\text{if $|\bsr_j|\le a_0/2$ and $|\bsr_k-(\ell,0)|\le a_0/2$},\\
0&\text{otherwise}.
\end{cases}
\en
selects configurations in which the particles $j$ and $k$ are near the origin and $(\ell,0)$, respectively.
Then we prove the following extension of the McBryan-Spencer inequality.

\prop{Theorem 2}%
For any $0<\beta<\infty$ and $a_1\le\ell\le L/2$, one has
\eq
|C_\ell(\beta)|\le\Bigl(\frac{\ell}{a_0}\Bigr)^{-\eta},
\lb{MS}
\en
with a positive constant $\eta$ that depends only on $\beta$,  $J_0$, $a_0$, and $a_1$.
(See \rlb{eta1} and \rlb{eta2}.)
There is no restriction on the particle number $N$.

Recall that the correlation function $C_\ell(\beta)$ should decay exponentially in $\ell$ when $\beta$ is sufficiently small.
(The exponential decay can be proved by invoking suitable expansion techniques.)
When the system becomes ordered (as in the two-dimensional Ising model at low temperatures) the correlation function at $h=0$ does not decay to zero, exhibiting long-range order.
The bound \rlb{MS} establishes that, at any nonzero temperature, the correlation function $C_\ell(\beta)$ decays at least by a power law, and hence never exhibits long-range order.
Note also that the bound is consistent with the expectation that the model undergoes a Berezinskii-Kosterlitz-Thouless type phase transition at a low temperature.

We shall prove \rlb{MS} by using the complex translation method of McBryan and Spencer \cite{MS}, along with its simplification by Picco \cite{P}.
The Hohenber-Mermin-Wagner type result \rlb{HMW} can be proved by using essentially the same techniques, as was first noted in \cite{TasakiBook}.

%%%%%%%
\para{Proof of \rlb{MS}}%
Let us write $\bsv_j=v_j(\cos\theta_j,\sin\theta_j)$ with $v_j\in[0,\infty)$ and $\theta_j\in[0,2\pi)$.
The velocity-dependent Hamiltonian \rlb{Hv} is written as
\eq
H_{\rm v}=-\sum_{j<k}J(\bsr_j,\bsr_k)\cos(\theta_j-\theta_k)-h\sum_j\cos\theta_j.
\en
It is also crucial to note that $H_{\rm p}$ is independent of $\theta_j$.

Let $h=0$.
We shall prove \rlb{MS} by setting $j=1$ and $k=2$ without losing generality.
Noting that the expectation value is invariant under $\theta_j\to-\theta_j$ for all $j$, we see that
$\bigl\langle\frac{\bsv_1}{|\bsv_1|}\cdot\frac{\bsv_2}{|\bsv_2|}\,\chi^{\ell}_{1,2}\bigr\rangle_{\beta,0}=\sbkt{\cos(\theta_1-\theta_2)\,\chi^{\ell}_{1,2}}_{\beta,0}=\sbkt{e^{i(\theta_1-\theta_2)}\,\chi^{\ell}_{1,2}}_{\beta,0}$.
We then note that
\eq
\sbkt{e^{i(\theta_1-\theta_2)}\,\chi^{\ell}_{1,2}}_{\beta,0}=Z_{\beta,0}^{-1}\int d\bsR\, dV\,e^{-\beta H_{\rm p}}\chi^{\ell}_{1,2}\,Y_{1,2}(\bsR),
\lb{ett}
\en
where $dV=\prod_{j=1}^Ndv_j\,v_j$ and  $\bsR=(\bsr_1,\ldots,\bsr_N)$.
We have defined
\eq
Y_{1,2}(\bsR)=\int d\Theta\,e^{i(\theta_1-\theta_2)+\sum_{j<k}\tJ_{j,k}\cos(\theta_j-\theta_k)},
\lb{Y12}
\en
where $d\Theta=\prod_{j=1}^Nd\theta_j$ and $\tJ_{j,k}=\beta J(\bsr_j,\bsr_k)$.
We understand that $Y_{1,2}(\bsR)$ is defined only for $\bsR$ such that $\sum_{j<k}u(\bsr_j,\bsr_k)<\infty$ and $\chi^{\ell}_{1,2}=1$.

For a fixed configuration $\bsR$, we choose $\ph_j\in\bbR$ for each $j$ and make the substitution $\theta_j\to\theta_j+i\ph_j$ in the integral \rlb{Y12}.
The integral is unchanged since contributions from the lateral contours cancel due to the $2\pi$ periodicity of the integrand \cite{SM}.
Recalling the identity $\cos(\theta+i\ph)=\cos\theta\cosh\ph-i\sin\theta\sinh\ph$, we see that
\eq
Y_{1,2}(\bsR)=e^{-\ph_1+\ph_2}\int d\Theta\,e^{iA+\sum_{j<k}\tJ_{j,k}\cos(\theta_j-\theta_k)\cosh(\ph_j-\ph_k)},
\lb{Y12B}
\en
where $A=\theta_1-\theta_2-\sum_{j<k}\tJ_{j,k}\sin(\theta_j-\theta_k)\sinh(\ph_j-\ph_k)$ is a real quantity.
We can then bound $|Y_{1,2}(\bsR)|$ as
\eqa
|Y_{1,2}(\bsR)|&\le e^{-\ph_1+\ph_2}\int d\Theta\,e^{\sum_{j<k}\tJ_{j,k}\cos(\theta_j-\theta_k)\cosh(\ph_j-\ph_k)}\nl
&\le e^{-\ph_1+\ph_2+\sum_{j<k}|\tJ_{j,k}|\,\{\cosh(\ph_j-\ph_k)-1\}}\,Y_0(\bsR)
\lb{Y12C}
\ena
with
\eq
Y_{0}(\bsR)=\int d\Theta\,e^{\sum_{j<k}\tJ_{j,k}\cos(\theta_j-\theta_k)}.
\lb{Y0}
\en
To get the second inequality in \rlb{Y12C}, we noted that 
$\cos\theta\,\cosh\ph=\cos\theta\{\cosh\ph-1\}+\cos\theta$, and used $|\cos\theta|\le1$.

Following \cite{P}, we set  $\ph_j=2\eta\log\{\ell/(|\bsr_j-\bsr_1|+a_0)\}$ if $|\bsr_j-\bsr_1|\le\ell-a_0$ and $\ph_j=0$ if $|\bsr_j-\bsr_1|\ge\ell-a_0$.
Note that $\ph_1=2\eta\log(\ell/a_0)>0$ and $\ph_2=0$ since $\chi^{\ell}_{1,2}=1$. 
We shall show below that the constant $\eta$ can be chosen so that the inequality
\eq
\sum_{j<k}|\tJ_{j,k}|\,\{\cosh(\ph_j-\ph_k)-1\}\le\eta\,\log\frac{\ell}{a_0}
\lb{main}
\en
holds.
Then \rlb{Y12C} implies $|Y_{1,2}(\bsR)|\le(\ell/a_0)^{-\eta}\,Y_0(\bsR)$.
Noting that
\eq
\sbkt{\chi^{\ell}_{1,2}}_{\beta,0}=Z_{\beta,0}^{-1}\int d\bsR\, dV\,e^{-\beta H_{\rm p}}\chi^{\ell}_{1,2}\,Y_{0}(\bsR),
\lb{echi}
\en
we see from \rlb{C} and \rlb{ett} that
\eq
|C_\ell(\beta)|=\frac{\bigl|\int d\bsR\int dV\,e^{-\beta H_{\rm p}}\chi^{\ell}_{1,2}\,Y_{1,2}(\bsR)\bigr|}{\int d\bsR\int dV\,e^{-\beta H_{\rm p}}\chi^{\ell}_{1,2}\,Y_{0}(\bsR)}\le\Bigl(\frac{\ell}{a_0}\Bigr)^{-\eta},
\en
which is the desired McBryan-Spencer bound \rlb{MS}.

%%%%%%%
\para{Proof of \rlb{HMW}}%
As was noted in section~4.4.3 of  \cite{TasakiBook}, the Hohenberg-Mermin-Wagner type theorem  \rlb{HMW} can also be proved by using the complex translation method.

Let $h>0$.
Again by symmetry one has $\bh{v_1^{\rm x}}=\bh{v_1^{\rm x}+iv_1^{\rm y}}=\bh{v_1e^{i\theta_1}}$.
As in \rlb{ett}, we have
\eq
\bh{v_1^{\rm x}}=Z_{\beta,h}^{-1}\int d\bsR\, dV\,v_1\,e^{-\beta H_{\rm p}}\,X_1(\bsR),
\lb{v11}
\en
with
\eq
X_1(\bsR)=\int d\Theta\,e^{i\theta_1+\sum_{j<k}\tJ_{j,k}\cos(\theta_j-\theta_k)+\beta h\sum_j\cos\theta_j}.
\en
By using the same complex translation, we can prove that
\eqa
|X_1(\bsR)|\le& e^{-\ph_1+\sum_{j<k}|\tJ_{j,k}|\,\{\cosh(\ph_j-\ph_k)-1\}+\beta h\sum_j(\cosh\ph_j-1)}\nl&\times X_0(\bsR),
\lb{X1B}
\ena
where
\eq
X_0(\bsR)=\int d\Theta\,e^{\sum_{j<k}\tJ_{j,k}\cos(\theta_j-\theta_k)+\beta h\sum_j\cos\theta_j}.
\en
Let us denote by $\Gamma_\eta(\ell)$ the maximum possible value of $\beta h\sum_j(\cosh\ph_j-1)$ for the same choice of $\ph_j$ as above.
We only need to know that $\Gamma_\eta(\ell)$ is finite and independent of the system size $L$ (provided that $\ell<L/2$).
Then \rlb{v11} and  \rlb{X1B}, with \rlb{main}, imply
\eq
\lim_{L\up\infty}\bigl|\bh{v_1^{\rm x}}\bigr|\le e^{-\eta\log(\ell/a_0)+\beta h\Gamma_\ell(\eta)}\,\lim_{L\up\infty}\bh{|\bsv_1|}
\en
for any $h>0$ and $\ell$.
This means that for any $h$ such that
\eq
0<h\le\frac{1}{2}\frac{1}{\beta\Gamma_\eta(\ell)}\,\eta\log\frac{\ell}{a_0},
\lb{h}
\en
one has
\eq
\lim_{L\up\infty}\bigl|\bh{v_1^{\rm x}}\bigr|\le e^{-(\eta/2)\log(\ell/a_0)}\,\lim_{L\up\infty}\bh{|\bsv_1|}.
\lb{Lv1x}
\en
By letting $\ell\up\infty$ while choosing $h$ in such a way that $h\dn0$ and \rlb{h} is always valid, we see that the left-hand side of \rlb{Lv1x} converges to zero as $h\dn0$.
Recalling that the particles are identical, we get the desired \rlb{HMW}.

%%%%%%%%%%%
\para{Proof of \rlb{main}}%
It remains to prove \rlb{main}.
We shall consider configurations in which particles are closely packed, and overestimate the sum in \rlb{main}.
We use the hard-core condition only in this estimate.

Let us set $\bsr_1=(0,0)$ for simplicity.
A rough estimate of the left-hand side of \rlb{main} is obtained by approximating $\cosh(\ph_j-\ph_k)-1$ by $(\ph_j-\ph_k)^2/2$, and by evaluating the sum as
\eqa
\sum_{j<k}(\ph_j-\ph_k)^2&\sim\int d^2\bsr|\nabla \ph(\bsr)|^2\sim\int_{|\bsr|\le\ell}\frac{d^2\bsr}{(|\bsr|+a_0)^2}
\nl&\sim\log(\ell/a_0).
\ena
Our task is to make this estimate into a rigorous bound.
It is tedious but is only technical.

For any $\bsr_j$, $\bsr_k$ with $|\bsr_j|\le\ell-a_0$ and $|\bsr_j-\bsr_k|\le a_1$, we see that
\eqa
|\ph_j-\ph_k|&\le2\eta\,\biggl|\log\frac{|\bsr_k|+a_0}{|\bsr_j|+a_0}\biggr|
\le2\eta\log\frac{|\bsr_j|+a_1+a_0}{|\bsr_j|+a_0}
\nl&=2\eta\log\biggl(1+\frac{a_1}{|\bsr_j|+a_0}\biggr)
\le\frac{2\eta\, a_1}{|\bsr_j|+a_0}.
\lb{pp}
\ena
For any $x_0>0$, it holds that $\cosh x-1\le(\cosh x_0-1)(x/x_0)^2$ for any $x$ such that $|x|\le x_0$ \cite{SM}.
Since we have $|\ph_j-\ph_k|\le2\eta a_1/a_0$ from \rlb{pp}, we find
\eqa
\cosh(\ph_j-\ph_k)-1&\le\zeta_0\Bigl\{\frac{a_0}{2\eta a_1}(\ph_j-\ph_k)\Bigr\}^2
\nl&\le\zeta_0\Bigl(\frac{a_0}{|\bsr_j|+a_0}\Bigr)^2,
\ena
where we again used $\rlb{pp}$ and set $\zeta_0=\cosh(2\eta a_1/a_0)-1$.
For a fixed $\bsr_j$ such that $|\bsr_j|\le\ell-a_0$, we thus have
\eqa
\sum_k|\tJ_{j,k}|\,&\{\cosh(\ph_j-\ph_k)-1\}
\nl&\le\Bigl(\frac{a_1+a_0}{a_0}\Bigr)^2 \beta J_0\zeta_0\Bigl(\frac{a_0}{|\bsr_j|+a_0}\Bigr)^2,
\ena
where we anticipated the worst case where the particle at $\bsr_j$ are closely surrounded by other particles within the radius $a_1$ and bounded the magnitude of the interaction by $J_0$.
We therefore find that
\eqa
\sum_{j<k}|\tJ_{j,k}|\,&\{\cosh(\ph_j-\ph_k)-1\}
\nl&\le\Bigl(\frac{a_1+a_0}{a_0}\Bigr)^2 \beta J_0\zeta_0\sum_{j=1}^n\Bigl(\frac{a_0}{|\bsr_j|+a_0}\Bigr)^2,
\lb{jkJ}
\ena
where $\bsr_1=(0,0)$, and other particles are closely packed in the sphere of radius $\ell-a_0$.
The sum is clearly bounded by an integral as
\eqa
\sum_{j=1}^n\Bigl(\frac{a_0}{|\bsr_j|+a_0}\Bigr)^2&\le \frac{C'}{(a_0)^2}\int_{|\bsr|\le\ell}d^2\bsr\Bigl(\frac{a_0}{|\bsr|+a_0}\Bigr)^2\nl&\le C\log\frac{\ell}{a_0},
\ena
where $C'$ and $C$ are numerical constants.
We thus get from \rlb{jkJ} that
\eqa
&\sum_{j<k}|\tJ_{j,k}|\,\{\cosh(\ph_j-\ph_k)-1\}
\nl&\le C \beta J_0\Bigl(\frac{a_1+a_0}{a_0}\Bigr)^2
\Bigl\{\cosh\Bigl(\frac{2\eta a_1}{a_0}\Bigr)-1\Bigr\}\log\frac{\ell}{a_0}
\ena
By choosing $\eta$ as  a unique positive solution of 
\eq
\eta=C \beta J_0\Bigl(\frac{a_1+a_0}{a_0}\Bigr)^2
\Bigl\{\cosh\Bigl(\frac{2\eta a_1}{a_0}\Bigr)-1\Bigr\},
\lb{eta1}
\en
we get the desired \rlb{main}.
Note that the solution always exists, and behaves as
\eq
\eta\simeq\Bigl\{2C\beta J_0\frac{(a_1+a_0)^2(a_1)^2}{(a_0)^4}\Bigr\}^{-1},
\lb{eta2}
\en
when $\beta$ is sufficiently large so that $\eta\ll a_0/a_1$ and $\cosh(2\eta a_1/a_0)-1\simeq(2\eta a_1/a_0)^2/2$.

\para{Discussion}%
We have proved that a class of two-dimensional particle systems with Vicsek-type ``exchange interaction'' never exhibits spontaneous breakdown of the rotational symmetry.
The conclusion is natural if one notices that, for each fixed particle configuration, the statistical behavior of the directions of the velocities is described by an effective XY spin system on a random network formed by particles.
This observation indeed played a key role in our proof.

Our results support the idea that the origin of the spontaneous symmetry breaking in the Vicsek and related models must be sought in the absence of the detailed balance condition \cite{TONER,TONER2,TONER3,TTR-REVIEW,DadhichiMaitraRamaswamy}.
It is an interesting challenge to rigorously understand what type of violation of detailed balance in microscopic dynamics leads to spontaneous symmetry breaking or, almost equivalently, to continuum dynamics as in \cite{TONER,TONER2,TONER3,TTR-REVIEW}.
See \cite{DadhichiKethapelliChajwaRamaswamyMaitra,ChenLeeToner} and references therein for promising directions.

Our theorems readily extend to a more general class of models with hard-core interactions, short-ranged exchange interactions, and global rotational symmetry for the velocities.
Rigorously speaking, our theorems do not cover the Hamiltonian flock models studied in \cite{HF1,HF2,HF3} since they do not satisfy the hard-core condition.
As is clear from our proof, however, the same conclusions should hold provided that particles do not exhibit pathological condensation.
It is likely that the apparent magnetic order observed numerically in \cite{HF2,HF3} is a manifestation of quasi long-range order characteristic in the Berezinskii-Kosterlitz-Thouless phase.
We also note that the theorems can easily be extended to models in which the exchange interaction in \rlb{Hv} is replaced by the nematic interaction, which is relevant to systems studied in \cite{NishiguchiNagaiChateSano2017,Tanida}.  See \cite{SM}.

\medskip
{\small
It is a pleasure to thank Shin-ichi Sasa for an inspiring discussion which motivated the present work.
I also thank Daiki Nishiguchi and Naoko Nakagawa for valuable discussions and comments, and  Hugues Chat\'e, Sriram Ramaswamy, and Masaki Sano for useful comments on the manuscript.
The present work was supported by JSPS Grants-in-Aid for Scientific Research no.~16H02211.
}

%%%%%%%%%%%%%%%%%%%%%%%%

%\end{document}

\clearpage

\begin{widetext}

\begin{center}
{\bf \Large Supplemental Material for ``Hohenberg-Mermin-Wagner type theorems for equilibrium models of flocking"}

\bigskip
Hal Tasaki
\end{center}

%\begin{quotation}
%In A of this supplemental material, we consider further examples of unique gapped ground states, and discuss the implication of our classification theorem.
%We also discuss the SU(3) AKLT type state in Fig.~\ref{f:su3} in detail.
%\end{quotation}

%%%%%%%%%%%%%%%%%%%%%%%%%%%%%%%%%%%%%%%
\bigskip\noindent
{\bf \large A. Invariance of the integral}
%%%%%%%%%%%%%%%%%%%%%%%%%%%%%%%%%%%%%%%
\setcounter{equation}{0}
\def\theequation{A.\arabic{equation}}
%%%%%%%%%%%%%%%%%%%%%%%%%%%%%%%%%%%%%%%

\medskip
Let us explain more carefully why the integral is unchanged by the substitution $\theta_j\to\theta_j+i\varphi_j$.
The invariance is indeed at the heart of the complex translation method of McBryan and Spencer \cite{MS}.

Let $f(z)$ be an analytic function of $z\in\bbC$.
Fix $\varphi\in\bbR$.
By modifying the integration path from $0\to 2\pi$ into $0\to i\varphi\to i\varphi+2\pi\to2\pi$, one has the identity
\eq
\int_0^{2\pi}d\theta\,f(\theta)=i\int_0^\varphi dy\,f(iy)+\int_0^{2\pi}d\theta\,f(\theta+i\varphi)-i\int_0^\varphi dy\,f(iy+2\pi).
\en
If the function satisfies the periodicity $f(z)=f(z+2\pi)$, this implies
\eq
\int_0^{2\pi}d\theta\,f(\theta)=\int_0^{2\pi}d\theta\,f(\theta+i\varphi).
\en

The desired invariance of the integral follows if we use this result for each $\theta_j$.

%%%%%%%%%%%%%%%%%%%%%%%%%%%%%%%%%%%%%%%
\bigskip\noindent
{\bf \large B. The bound for $\cosh x-1$}
%%%%%%%%%%%%%%%%%%%%%%%%%%%%%%%%%%%%%%%
\setcounter{equation}{0}
\def\theequation{B.\arabic{equation}}
%%%%%%%%%%%%%%%%%%%%%%%%%%%%%%%%%%%%%%%

\medskip
Let us prove, for completeness, the technical upper bound for $\cosh x-1$ that appears below (25).

We shall show that
\eq
f(x):=\frac{\cosh x-1}{x^2}
\en
is nondecreasing in $x$ for $x\ge0$.
Then it follows for any $0\le x\le x_0$ that 
\eq
\frac{\cosh x-1}{x^2}\le\frac{\cosh x_0-1}{(x_0)^2},
\en
which gives the desired bound.

To show the claim for $f(x)$, we note that
\eq
f'(x)=\frac{g(x)}{x^3}
\en
with
\eq
g(x)=2(1-\cosh x)+x\,\sinh x.
\en
The derivatives of $g(x)$ are found to be
\eq
g'(x)=x\,\cosh x-\sinh x,\quad g''(x)=x\,\sinh x.
\en
Since $g''(x)\ge0$ for $x\ge0$ and $g'(0)=0$, we see that $g'(x)\ge0$ for $x\ge0$.
Then since $g(0)=0$, we see that $g(x)\ge0$ for $x\ge0$.

%%%%%%%%%%%%%%%%%%%%%%%%%%%%%%%%%%%%%%%
\bigskip\noindent
{\bf \large C. Nematic intereaction}
%%%%%%%%%%%%%%%%%%%%%%%%%%%%%%%%%%%%%%%
\setcounter{equation}{0}
\def\theequation{C.\arabic{equation}}
%%%%%%%%%%%%%%%%%%%%%%%%%%%%%%%%%%%%%%%

\medskip

Let us discuss the extensions of our results to models with nematic interaction.

We here consider a model obtained by replacing the 
Hamiltonian $H_{\rm v}$ in \rlb{Hv} with
\eq
\tilde{H}_{\rm v}=-\mathop{\sum_{j,k=1}^N}_{(j<k)}J(\bsr_j,\bsr_k)\Bigl(\frac{\bsv_j}{|\bsv_j|}\cdot\frac{\bsv_k}{|\bsv_k|}\Bigr)^2
-h\sum_{j=1}^N\frac{(v_j^{\rm x})^2}{|\bsv_j|^2}.
\lb{tHv}
\en
Clearly the nematic interaction $-J(\bsr_j,\bsr_k)(\frac{\bsv_j}{|\bsv_j|}\cdot\frac{\bsv_k}{|\bsv_k|})^2$ favors states in which the velocities of two nearby particles are parallel or antiparallel.
Similarly the symmetry breaking field  $-h\frac{(v_j^{\rm x})^2}{|\bsv_j|^2}$ tends to align the velocities in the positive or negative x-direction.
Note that the new Hamiltonian \rlb{tHv} is written in terms of the angle variables as
\eq
\tilde{H}_{\rm v}=-\mathop{\sum_{j,k=1}^N}_{(j<k)}J(\bsr_j,\bsr_k)\bigl\{\cos(\theta_j-\theta_k)\bigr\}^2
-h\sum_{j=1}^N\bigl\{\cos(\theta_j)\bigr\}^2.
\lb{tHv2}
\en

We then study the equilibrium state of the Hamiltonian $\tilde{H}=H_{\rm p}+\tilde{H}_{\rm v}$.
The following is a direct analogue of Theorem~1, the extended Hohenberg-Mermin-Wagner theorem.

\medskip\noindent
{\bf Theorem 3:}
For any $0<\beta<\infty$ one has
\eq
\lim_{h\dn0}\lim_{L\up\infty}\frac{1}{N}\sum_{j=1}^N\sbkt{e^{2i\theta_j}}_{\beta,h}=0,
\lb{HMWt}
\en
where the particle number $N$ may depend in an arbitrary manner on the system size $L$.

\medskip

The nematic correlation between two particles is most conveniently measured by the conditional correlation fucntion
\eq
\tilde{C}_\ell(\beta)=
\frac{\bigl\langle e^{2i(\theta_j-\theta_k)}\,\chi^{\ell}_{j,k}\bigr\rangle_{\beta,0}}{\sbkt{\chi^{\ell}_{j,k}}_{\beta,0}},
\lb{tC}
\en
where it should be noted that
\eq
e^{2i(\theta_j-\theta_k)}=\biggl(\frac{v_j^{\rm x}+iv_j^{\rm y}}{|\bsv_j|}\frac{v_k^{\rm x}-iv_k^{\rm y}}{|\bsv_k|}\biggr)^2.
\en
We have the following analogue of Theorem~2, the McBryan-Spencer bound.

\medskip\noindent
{\bf Theorem 4:}
For any $0<\beta<\infty$ and $a_1\le\ell\le L/2$, one has
\eq
|\tilde{C}_\ell(\beta)|\le\Bigl(\frac{\ell}{a_0}\Bigr)^{-\tilde{\eta}},
\lb{MSt}
\en
with a positive constant $\tilde{\eta}$ that depends only on $\beta$,  $J_0$, $a_0$, and $a_1$.
There is no restriction on the particle number $N$.

\newcommand{\Dt}{\mathit{\Delta}\theta}
\newcommand{\Dp}{\mathit{\Delta}\varphi}

\medskip

The proofs are essentially the same as those in the main text.
The only crucial difference is the change in the Boltzmann factor, which can be treated as follows.
Let $\Dt:=\theta_j-\theta_k$ and $\Dp=\varphi_j-\varphi_k$.
Then we now have the term $\tilde{J}_{j,k}\{\cos\Dt\}^2$ in the Boltzmann factor of the quantity corresponding to (10).
After the complex translation $\theta_j\to\theta_j+i\varphi_j$, this becomes $\tilde{J}_{j,k}\{\cos(\Dt+i\Dp)\}^2$.
We then note  that
\eqa
\{\cos(\Dt+i\Dp)\}^2&=\{\cos\Dt\,\cosh\Dp-i\,\sin\Dt\,\sinh\Dp\}^2
\nl&=(\cos\Dt)^2(\cosh\Dp)^2-(\sin\Dt)^2(\sinh\Dp)^2+(\text{pure imaginary quantity}).
\ena
The pure imaginary term can be dealt with as in the main text.
We then bound the real part as
\eqa
\tilde{J}_{j,k}\bigl\{(\cos\Dt)^2&(\cosh\Dp)^2-(\sin\Dt)^2(\sinh\Dp)^2\bigr\}
\nl&=\tilde{J}_{j,k}\bigl\{(\cos\Dt)^2+(\cos\Dt)^2\{(\cosh\Dp)^2-1\}-(\sin\Dt)^2(\sinh\Dp)^2\bigr\}
\nl&\le \tilde{J}_{j,k}(\cos\Dt)^2+|\tilde{J}_{j,k}|\bigl\{(\cosh\Dp)^2-1+(\sinh\Dp)^2\bigr\}.
\ena
We see that this is in good shape since $(\cosh\Dp)^2-1+(\sinh\Dp)^2\simeq2(\Dp)^2$ for small $\Dp$.
Rigorous upper bound can be constructed as in the original proof.

\end{widetext}

\end{document}